\documentstyle[dina4p,11pt]{article}
\def\E{{\cal E}}
\def\o{\omega}
\def\r{\rho}
\def\f{\frac}

\def\s{\sqrt}

\def\be{\begin{equation}}
\def\ee{\end{equation}}
\def\la{\label}
\def\i{\infty}
\def\x{{\vert x \vert}}
\def\Lh{{\frac{1}{2} L}}

\begin{document}
\newtheorem{theorem}{Theorem}
\begin{center}
\LARGE{A Periodic Analog of the Schwarzschild Solution}
\vskip2.0cm
\large{D.Korotkin\footnote{Supported by Alexander von Humboldt
      Foundation.}} and \large{H.Nicolai}
\vskip1.0cm
II. Institute for Theoretical Physics, Hamburg University, \\
Luruper Chaussee 149,  Hamburg 22761, Germany
\end{center}
\vskip2.5cm
{\bf ABSTRACT.} We construct a new exact solution of Einstein's
equations in vacuo in terms of Weyl canonical coordinates. This
solution may be interpreted as a black hole in a space-time which is
periodic in one direction and which behaves asymptotically like the
Kasner solution with Kasner index equal to
$4M L^{-1}$, where $L$ is the period and $M$ is the mass of the
black hole. Outside the horizon, the solution is free of singularities
and approaches the Schwarzschild solution as $L \rightarrow \infty$.
\vskip2.0cm
In this article, we present a new exact solution of Einstein's
equations in vacuo (for a comprehensive review of exact solutions,
see \cite{book}). This solution constitutes a periodic analog of the
Schwarzschild solution \cite{S}, and is likewise free of singularities.
Asymptotically, it behaves like the Kasner solution.
Before describing the new solution, we would like to remind the
reader of the link between meromorphic functions
on the Riemann sphere ${\bf C}P^1$ and on the torus
$T={\bf C}/\{L_1,L_2\}$ ($L_1$ and $L_2$ are the two periods on the
lattice defining $T$). Given some meromorphic function $f_0(\xi)$,
$\xi \in {\bf C}$, we can consider the formal expression
\be
f(\xi)=\sum_{m, n =-\i}^{\i} \Big\{f_0(\xi +m L_1 +n L_2)+ a_{mn}\Big\}
\la{f}\ee
Provided we can choose the constants $a_{mn}$ in such
a way that this series converges, the function $f(\xi)$ is
meromorphic on the torus $T$ with the same number and
positions of poles as the function $f_0 (\xi)$ on ${\bf C}P^1$.
In this case, $f(\xi)$ is a doubly periodic analog
of the original function $f_0$. The simplest example of a function
$f_0(\xi)$ admitting an appropriate choice of convergence generating
constants $a_{mn}$ is $ f_0(\xi) = \xi^{-2}$
with $a_{mn} = - (mL_1 +nL_2)^{-2}$ for $(m,n) \neq (0,0)$ and
$a_{00}=0$, leading to the well-known Weierstrass $\wp$-function.
(For $ f_0(\xi) = \xi^{-1}$, on the other hand,
no choice of $a_{mn}$ will render the series (\ref{f}) convergent
since this would imply the existence of a meromorphic function on the
torus having only one pole in contradiction to the Riemann-Roch
theorem.)

The same procedure works, of course, for any {\it real}
harmonic function $\o(\xi,\bar{\xi})$ which can be represented as
$\o = {\rm Re} f(\xi)$ for some (locally) holomorphic function
$f$. For $\o (\xi,\bar{\xi})$ we have the Laplace equation
\be
\o_{\xi\bar{\xi}} =0
\la{LE}\ee
Real doubly periodic solutions of (\ref{LE}) may be constructed
starting from an arbitrary solution $\o_0(\xi, \bar{\xi})$ of (\ref{LE})
if we can ensure the convergence of the series
\be
\sum_{m,n=-\i}^{\i} \Big\{ \o_0(\xi +mL_1 +nL_2) +b_{mn} \Big\}
\la{op}\ee
by appropriate choice of the real constants $b_{mn}$.
We emphasize that this construction rests on two features of the
Laplace equation (\ref{LE}), namely (i) its linearity and (ii)
on its invariance with respect to arbitrary
translations $\xi \rightarrow \xi + L$, $L\in {\bf C}$.

Now consider the Euler-Darboux equation
\be
\o_{\xi\bar{\xi}} - \f{\o_{\xi} -\o_{\bar{\xi}}}{2(\xi -\bar{\xi})}
=0 \;,\;\;\;\;\o(\xi,\bar{\xi})\in {\bf R}
\la{ED}\ee
or, in terms of real coordinates $(x,\rho )$ (where $\xi = x+i\rho$),
\[ \o_{xx} +\f{1}{\r}\o_{\rho} +\o_{\rho\rho} =0 \]
This equation satisfies the same properties as (2) with one important
modification: it is invariant with respect to {\it real}
translations $\xi\rightarrow\xi + L$ only (i.e. $L\in {\bf R}$).
Thus there are no doubly periodic solutions of (\ref{ED}), but we can
still try to apply a similar scheme to obtain solutions of
(\ref{ED}) which are periodic in the $x$-direction. Namely, let
$\o_0 (x,\r)$ be some solution of (\ref{ED}). In analogy with
(\ref{op}) we consider the expression
\be
\o(x,\rho)=\sum_{n=-\i}^{\i}\Big\{\o_0(x+nL,\rho) + a_n\Big\}
\la{ps}\ee
where $a_n$ are constants to be chosen in such a way that the
series (\ref{ps}) becomes convergent. It is important that these
coefficients do not depend on $(x,\rho)$ since otherwise the sum could
not possibly satisfy the original equation (\ref{ED}). If convergence
can be achieved, the function (\ref{ps}) describes a solution of
(\ref{ED}) with period $L$.

In order to exploit these observations for the construction of new
solutions of Einstein's equations, we start from a
stationary axisymmetric space-time with the metric
\be
ds^2 =f^{-1}\big[e^{2k}(dx^2 +d\r^2) +\r^2 d\phi^2\big] +f(dt +Ad\phi)^2
\la{m}\ee
where $(x,\rho)$ are Weyl canonical coordinates (here $\rho \geq 0$
is the radial coordinate). The functions $f(x,\rho ) \,,\,
k(x,\rho)$ and $A(x,\rho)$ are related to the complex-valued
Ernst potential $\E(x,\rho)$ \cite{Ernst} as follows:
\be
f={\rm Re}\E\;\;\;\;\;\;A_{\xi}=2\rho\f{(\E-\bar{\E})_{\xi}}
{(\E+\bar{\E})^2} \;\;\;\;\;\;
k_{\xi}= 2i\rho\f{\E_{\xi}\bar{\E}_{\xi}}{(\E+\bar{\E})^2}
\la{c}\ee
where again $\xi=x+ i\rho$. In terms of $\E(x,\rho)$, the Einstein
equations may be written in the form
\be
(\E+\bar{\E})(\E_{xx}+\f{1}{\rho}\E_{\rho}+\E_{\rho\rho})
=2(\E_x^2 +\E_{\rho}^2)
\la{ee}\ee
If $\E\in {\bf R}$, the coefficient $A$ in (\ref{m}) vanishes, and the
metric (\ref{m}) becomes static, i.e. describes a space-time without
rotation. In this case, the Ernst equation (\ref{ee}) can be
linearized by defining
\[ \o =\log \E \]
and is thereby reduced to the Euler-Darboux equation
(\ref{ED}). The metric (\ref{m}) becomes
\be
ds^2 =e^{-\o}\big[e^{2k}(dx^2 +d\r^2) +\r^2 d\phi^2\big] + e^{\o} dt^2
\la{m1}\ee
where the conformal factor is determined from the first order equation
\[ k_{\xi} =\f{i\rho}{2}(\o_{\xi})^2 \]
or, equivalently,
\be
k_{\rho}=\f{\rho}{4} \big( \o_{\rho}^2 -\o_x^2\big)\;\;\;\;\;\;\;
k_x=\f{\rho}{2}\o_x\o_{\rho}
\la{cs}\ee
We can now construct $x$-periodic analogs of known static
solutions by means of the procedure outlined above. Note that in order
to obtain a truly periodic metric (\ref{m1}) we must also verify the
periodicity of the function $k(x,\rho)$ defined by (\ref{cs}).

The following theorem shows that our method of construction
is really quite general.

\begin{theorem}
Let $\o_0(x,\rho)$ be any solution of the Euler-Darboux equation
corresponding to an asymptotically flat metric (\ref{m1}), i.e.
\be
\o_0(x,\rho)=\f{\beta}{r} +O(r^{-2})\;\;\;\;\;{\rm as}\;\;\;\;\;
r\rightarrow\i
\la{af}\ee
where $r=\sqrt{x^2 +\rho^2}$; $M=-\f{1}{2} \beta$ is the mass. Let
\be
a_n = -\f{\beta}{L|n|}\;\;,\;\;\;n\neq0\;\;,\;\;a_0=0
\la{an}\ee
Then series (\ref{ps}) is convergent for all $(x,\rho)$ except
the points $(x_0+nL,\rho_0)$, where the function
$\o_0(x,\rho)$ is singular ($n\in {\bf Z}$), and defines a periodic
function with period $L$.
\end{theorem}

{\it Proof:} For large $n$ we have
\[ \o_0(x+nL,\rho) + a_n = \beta\Big(\f{1}{\sqrt{(x+nL)^2+\rho^2}} -
\f{1}{L|n|}\Big) + O\big(\f{1}{n^2}\big) = O\big(\f{1}{n^2}\big) \]
by (\ref{af}), and therefore the series (\ref{ps}) converges if
$(x+nL,\rho)$ does not coincide with a singular point of $\o_0(x,\rho)$
for any $n$. $\Box$

So starting from an arbitrary static asymptotically flat solution
we can construct its $x$-periodic analog. In the remainder we will,
however, restrict attention to the Schwarzschild solution,
which is characterized by the Ernst potential
\be
\o_0=\log\E_0 \;\;\;\;\;\;\;
\E_0 (x,\rho ) =\f{\s{(x-M)^2+\rho^2} + \s{(x+M)^2+\rho^2} -2M}
{\s{(x-M)^2+\rho^2} + \s{(x+M)^2+\rho^2} +2M}
\la{S}\ee
where $M \in {\bf R}$ is an arbitrary positive constant (the mass
of the black hole). Here, all square roots are taken to be positive;
this means that we do not consider (\ref{S}) inside the event horizon,
which coincides with the segment $\rho=0\,,\,x\in [-M,M]$.
The coefficient $\beta$ in (\ref{af}) is therefore
equal to $-2M$. Thus the periodic analog of the
Schwarzschild solution (\ref{S}) has the following form:
\be
\E(x,\rho)=\E_0(x,\rho)\prod_{n=1}^{\i}\E_0(x+nL,\rho)\E_0(x-nL,\rho)
\exp\Big( \f{4M}{nL} \Big)
\la{pS}\ee
Obviously,
\[ \E(x+L,\rho)=\E(x,\rho) \]
In the sequel we will assume $\f{1}{2} L > M$ (for $L\leq 2M$,
the interpretation of the solution is not clear
as the horizon overlaps with itself, and the Ernst potential
vanishes on the symmetry axis). Convergence
of the infinite product (\ref{pS}) is equivalent to convergence of the
series (\ref{ps}) and thus guaranteed by Theorem 1 (for non-negative
values of the square roots in (\ref{S})). Consequently, the
solution (\ref{pS}) is a periodic function on the upper half plane
$\rho \geq 0$ with ``fundamental region"  $\cal F$ defined by
$\rho \geq 0\,,\,-\f{1}{2} L\leq x \leq\f{1}{2} L$.

\begin{theorem}
The function $\E(x,\rho)$ defined by (\ref{pS}) is smooth
everywhere on $\cal F$ away from the points $x=\pm M\,,\,\rho =0$.
Moreover, it is non-zero everywhere except on the horizon
(i.e. $\rho =0 \,,\, \x \leq M$).
\end{theorem}
{\it Proof:} These properties are a consequence of analogous
properties of the Schwarzschild solution and the convergence
of (\ref{ps}).
$\Box$

As mentioned above we must also check the periodicity of $k(x,\rho )$.

\begin{theorem}
Let $L > 2M$. Then the function $k(x,\rho)$ corresponding to the
Ernst potential (\ref{cs}) is periodic in $x$ with period $L$, i.e.
\[ k(x+L,\rho)= k(x,\rho) \]
\end{theorem}
{\it Proof:} It is sufficient to show that
\[ \int_{-L/2}^{L/2} k_x dx = 0 \]
where the derivative $k_x$ is to be evaluated by means
of (\ref{cs}). Consider the integral
\be
\int_{l} \Big\{\f{\rho}{4}(\o_{\rho}^2 -\o_x^2)d\rho +\f{\rho}{2}
\o_x\o_{\rho} dx\Big\}
\la{ci}\ee
where the closed contour $l$ is chosen according to Fig.1. This
integral vanishes because the function $\o$ obeys (\ref{ED}) and is
smooth everywhere inside of $l$. The integrals along the edges
$[(-\Lh,0),\;(-\Lh,\rho)]$ and $[(\Lh,\rho),\;(\Lh,0)]$ cancel
due to the periodicity of $\o(x,\rho)$. Owing to the presence of the
factor $\rho$ in (\ref{ci}) the contribution of the interval
$[(\Lh,0),\;(-\Lh,0)]$ reduces to a sum of contributions of two small
rectangular paths around the points $x=-M$ and $M$ (cf. Fig.1.),
where the derivatives $\o_x$ and $\o_{\rho}$ become singular. These
contributions cancel by virtue of the symmetry
\[ \o(-x,\rho) =\o(x,\rho) \]
inherited by solution $\E$ from $\E_0$. So the integral along the
contour $[(-\Lh,\rho),\;(\Lh,\rho)]$ also vanishes and we get
$k(-\Lh,\rho)=k(\Lh,\rho)$. $\Box$

Hence the metric (\ref{m1}) corresponding to the periodic
solution (\ref{pS}) is also periodic. Furthermore, we have

\begin{theorem}
The asymptotic behavior of the Ernst potential (\ref{pS}) is given by
\be
\E=C\rho^{4M/L} \Big( 1+o(1)\Big)\;\;\;{\rm as}\;\;\;\rho\rightarrow\i
\la{ass}\ee
where $C$ is some constant.
\end{theorem}
{\it Proof:} The function $\o=\log\E$ is defined by
\[ \o(x,\rho)=\o_0(x,\rho) + \sum_{n=1}^{\i}\Big[\o_0(x+nL,\rho) +
\o_0(x-nL,\rho) +\f{4M}{nL}\Big] \]
Substituting the explicit expression for $\o_0(x,\rho)$ (\ref{S})
and differentiating with respect to $\rho^2$, we obtain
\[ \frac{\partial\o}{\partial(\rho^2)} (x,\rho)=
\sum_{n=-\i}^{\i} \f{2M \big[s_1(n) + s_2(n) \big]}{\big[
   s_1(n)+s_2(n)+2M\big]\big[s_1(n)+s_2(n)-2M\big]
s_1(n) s_2(n)} \]
where $s_1(n)=\s{(x+nL+M)^2+\rho^2}$ and
$s_2(n)=\s{(x+nL-M)^2 +\rho^2}$.
The leading term in this series for large $\rho$ can be estimated
by approximating the sum by an integral; it is given by
\[ \sum_{n=-\i}^{\i}\f{M}{((x+nL)^2 +\rho^2)^{3/2}} =
       \f{2M}{L}\f{1}{\rho^2} \Big( 1+o(1) \Big) \]
Thus,
\[ \o = \f{2M}{L}\log\rho^2 + O(1)\;\;\;\;\;\;{\rm as}
\;\;\;\;\rho\rightarrow \i \]
and $\E = C\rho^{4M/L} (1+o(1))$ for some constant $C$. $\Box$

Hence, as $\rho \rightarrow \infty$, the metric
(\ref{m1}) tends to the Kasner solution
\be
ds^2= \tilde C \rho^{\f{\alpha^2}{2} -\alpha} (dx^2 +d\rho^2)
   + C^{-1} \rho^{2-\alpha}
d\phi^2 - C \rho^{\alpha} dt^2
\la{as}\ee
where $\tilde C$ is another constant of integration and
the Kasner parameter $\alpha$ is related to the period $L$ by
$\alpha=4M L^{-1}$, so that $0\leq \alpha <2$ with our assumption
on the range of $M$.

The new solution has a compact event horizon coinciding with the
segment $\rho=0\,,\,-M\leq x\leq M$. Outside the horizon it is
everywhere non-singular, including the segment of the symmetry axis
outside the horizon. Using the standard product representation
for the $\Gamma$-function, we find
\be
\E(x, \rho =0)= \exp \Big( \f{4\gamma M}{L} \Big) \,
\f{ \Gamma \Big( \f{\x +M}{L} \Big)
    \Gamma \Big( 1-\f{\x -M}{L} \Big)}
  { \Gamma \Big( \f{\x -M}{L} \Big)
    \Gamma \Big( 1-\f{\x +M}{L} \Big)}
\ee
for $M \leq \x \leq \f{1}{2} L $ ($\gamma$ is the Euler-Mascheroni
constant), and $\E \equiv 0$ for $\rho =0$ and $\vert x \vert \leq M$.
As a consequence of the reflection and translation symmetry, the
free integration constant in equation (\ref{cs})
may be chosen in such a manner that conical singularities on the
part of the symmetry axis outside the horizon are avoided
(this requirement fixes the constant $\tilde C$ in (\ref{as})).
In the limit $L\rightarrow\i$ the solution obviously tends (pointwise)
to the ordinary Schwarzschild solution. The leading term in the
asymptotic expansion then approaches the flat metric, as it should be.
Alternatively, one can regard the new solution as describing an
infinite chain of black holes spaced at a distance $L$. At first sight,
it seems remarkable that this configuration does not require conical
singularities on the axis between adjacent black holes for
stability. Rather, it appears to be stabilized by its symmetry under
reflections and translations and the presence of infinitely many
black holes ``on each side". However, this also indicates instability
under non-periodic perturbations, which makes this interpretation
somewhat less attractive.

In summary, our new solution can be interpreted as a natural analog
of the Schwarzschild solution in an $x$-periodic Kasner universe.
The existence of this solution does not in any way contradict
well-known theorems on the uniqueness of black hole solutions \cite{bh}
as these theorems only apply to asymptotically flat space-times.
Note that the product (\ref{pS}) is also convergent inside the event
horizon, where $\rho$ becomes imaginary if all square roots in (\ref{S})
are taken to be positive. On the second sheet of the ordinary
Schwarzschild solution, where the square roots become negative and the
mass is also negative, the product (\ref{pS}) diverges, i.e.
$\E\rightarrow\i$, so the second sheet is, in fact, absent.
It would be desirable to study the behaviour of our solution on and
inside the event horizon and to compare it with the ordinary
Schwarzschild solution there. Furthermore, the Kasner solution
(\ref{as}) represents only the main term in the asymptotic
expansion of our solution. Clearly, it would be interesting to
identify the next order term and to define some analog of mass
in the periodic case.
Finally, the construction of periodic analogs of the Kerr solution
\cite{K} will be more complicated as the Ernst equation can no longer
be linearized in that case. In addition, one must presumably take
into account infinite-soliton Ernst potentials.

\vskip 14cm
\begin{center}
{\bf Figure 1:} Integration contour used in the proof of Theorem 3.
\end{center}

\begin{thebibliography}{99}
\bibitem{book} D. Kramer, H. Stephani, E. Herlt and M. MacCallum,
{\it Exact Solutions of Einstein's Field Equations}
(Cambridge University Press, 1980)
\bibitem{S} K.Schwarzschild, Sitzungsber. Preuss. Akad. Wiss.,
            Kl. Math.-Phys. Tech., 189 (1916)
\bibitem{Ernst} F. Ernst, Phys. Rev. {\bf 167}, 1175 (1968)
\bibitem{bh} S. Chandrasekhar, {\it The Mathematical Theory of Black
     Holes} (Oxford University Press, 1983)
\bibitem{K} R.P.Kerr, Rhys. Rev. Lett. {\bf 11}, 237 (1963)
\end{thebibliography}
\end{document}